\documentclass[twocolumn,showpacs,reprint,superscriptaddress,prl,floatfix]{revtex4-1}

\usepackage{amsmath, amsthm, amsfonts}    
\usepackage{amssymb}
\usepackage{bm}            
\usepackage{color}

\usepackage{soul} 
\usepackage{mathtools} 
\usepackage{tikz} 
\usepackage{cancel} 

\pdfoutput=1
\usepackage[unicode=true,pdfusetitle,
 bookmarks=true,bookmarksnumbered=false,bookmarksopen=false,
 breaklinks=true,pdfborder={0 0 0},backref=false,colorlinks=true,citecolor=blue]{hyperref}

\usepackage{graphicx}   
\usepackage[caption=false]{subfig}       
\usepackage{etoolbox} 
\usepackage{svg} 

\usepackage[nodisplayskipstretch]{setspace}
\usepackage{float}
\usepackage{epstopdf}

\makeatother
   
\def\be{\begin{eqnarray}}
\def\ee{\end{eqnarray}}

\begin{document}

\title{Asymmetry and spin-orbit coupling  of light scattered from subwavelength  particles}

\author{Jorge Olmos-Trigo}
\affiliation{Donostia
International Physics Center (DIPC), 
20018 Donostia-San Sebastian, Spain}

\author{Cristina Sanz-Fern\'andez}
\affiliation{
Centro de F\'{i}sica de Materiales (CFM-MPC), Centro Mixto CSIC-UPV/EHU,  20018 Donostia-San Sebasti\'{a}n,  Spain}

\author{F. Sebasti\'an Bergeret}
\affiliation{Donostia
International Physics Center (DIPC), 
20018
Donostia-San Sebastian, Spain}
\affiliation{
Centro de F\'{i}sica de Materiales (CFM-MPC), Centro Mixto CSIC-UPV/EHU,  20018 Donostia-San Sebasti\'{a}n,  Spain}

\author{Juan Jos\'e S\'aenz}
\affiliation{Donostia
International Physics Center (DIPC), 
20018
Donostia-San Sebastian, Spain}
\affiliation{IKERBASQUE, Basque Foundation for Science, 48013 Bilbao, Spain}

\date{\today}

\begin{abstract}
Light scattering and spin-orbit angular momentum coupling phenomena from subwavelength objects, with electric and magnetic dipolar responses, are receiving an increasing interest.   Under illumination by circularly polarized light,  spin-orbit coupling effects  have been shown to  lead to significant shifts between the measured and actual position of particles.
 Here we show that the remarkable angular dependence of these ``optical mirages'' and those of the 
 intensity, degree of circular polarization (DoCP), and  spin and orbital angular momentum of scattered photons,  are all linked  and fully determined  by  the dimensionless  ``asymmetry parameter'' $g$, being independent of  the specific optical properties of the scatterer. Interestingly,  for $g \neq 0$  the maxima of the optical mirage and angular momentum exchange take place at different scattering angles.  In addition we  show that the $g$ parameter is exactly half of   the DoCP  at a right-angle scattering. This finding opens the possibility to  infer the whole  angular  properties of the scattered fields by a single far-field polarization measurement. 
 \end{abstract}
\maketitle

The interference between electric and magnetic dipolar fields scattered from
high refractive index (HRI) subwavelength particles  is known to lead to strong asymmetric intensity distributions \cite{geffrin2012magnetic}, electric-magnetic radiation pressure effects, \cite{nieto2010optical} and  other interesting phenomena with novel physical effects and applications
\cite{kuznetsov2016optically}. In addition to energy and linear momentum, a light wave carries angular momentum (AM) \cite{allen2003optical} that can be split into spin (SAM) and orbital angular momentum ({OAM}) \cite{he1995direct,simpson1997mechanical,crichton2000measurable}. Light scattering  may couple these two components of the AM, via the spin-orbit interaction (SOI) and modify the contributions of SAM and OAM. This phenomena  has attracted a great deal of attention in  recent years \cite{berry2005orbital,bliokh2015spin}.

Among all the intriguing effects originated by the  {SOI}, perhaps the most interesting  is the appearance of the {optical mirage}, {\it i.e.},  an apparent transversal displacement of a target after scattering \cite{onoda2004hall,bliokh2006conservation}. This apparent shift, induced by the AM exchange per photon, has been predicted and experimentally proved in very different  situations. These include circularly polarizedlight impinging a dielectric surface \cite{hosten2008observation} or  a single electric  dipolar particle \cite{schwartz2006conservation,arnoldus2008subwavelength}, where the absolute values of the optical mirage  are limited to subwalengths scales. It has been recently demonstrated that this dipolar limit can be surpassed by illumination with elliptically polarized  light \citep{araneda2018wavelength}. Analogously, enhanced optical mirage values (reaching tens of wavelengths) were obtained for resonant Mie scatters, where higher multipoles are needed \citep{haefner2009spin}. 
In particular, it has been shown that a high refractive index (HRI)  Si-sphere with electric  and magnetic dipolar response \cite{evlyukhin2010optical,Garcia-Etxarri2011}, can lead to a diverging optical mirage at backscattering \cite{Olmos2018SOI,gao2018enhanced,olmos2018enhanced},  when the helicity is preserved \cite{zambrana2013duality}.
These findings may  give ground for  the conjecture that any optical property  related to the  electric and magnetic polarizabilities, such as absorption, particle size or refractive index, may modify the helicity pattern and hence the optical mirage.

In this Letter, we demonstrate that the degree of circular polarization (DoCP) or, equivalently, the helicity density, $\Lambda_\theta$ \citep{bohren2008absorption}, depends indeed on the optical properties only through the asymmetry parameter  $g$ in the dipolar regime \cite{gomez2012negative}.  We demonstrate that  from the DoCP  measurement in the far field limit (FF) at a single scattering angle, one can extract the  full information about  other optical parameters, such as $g$, the re-distribution of AM and the optical mirage. 
Interestingly, it follows from our study  that for a non-zero $g$, the maximum exchange of AM  and the  maximum optical mirage value, do not occur at the same scattering angle.  At  these maxima, the polarization of light is not lineal but elliptical with a $z$-component of the OAM larger than the total AM, in striking contrast with the pure electric (or magnetic) case $g=0$ \cite{schwartz2006conservation,arnoldus2008subwavelength}. 

Specifically, we consider a dielectric sphere of radius $a$ with an arbitrary  permittivity $\epsilon_{\rm{p}}$ and refractive index $m^2_{\rm{p}} = \epsilon_{\rm{p}}$, which is embedded in an otherwise homogeneous medium  with constant and real relative dielectric permittivity $\epsilon_{\rm{h}}$ and  refractive index $m^2_{\rm{h}} = \epsilon_{\rm{h}}$. The geometry of the scattering is sketched in FIG. 1, where a circularly polarized plane wave, with wavenumber $k$ ($k =  m_h 2 \pi / \lambda_0$, being  $ \lambda_0$ the  wavelength in vacuum) and  well-defined helicity $\sigma = \pm 1$, is incoming along the $z$-axis.
\begin{figure}
\centering
\fbox{
\includegraphics[width = 0.9\columnwidth]{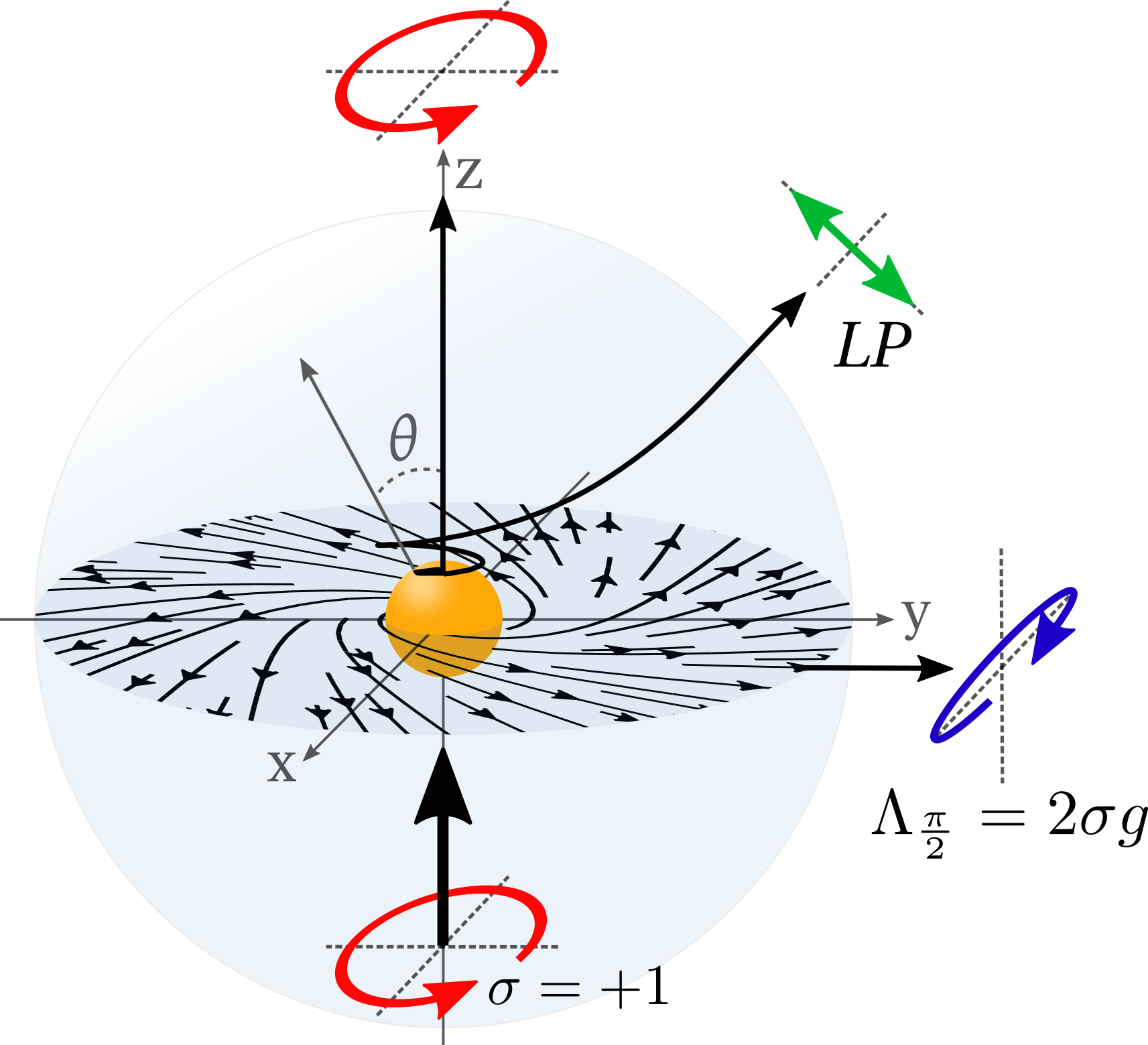}
}
\caption{Sketch of the system: plane wave with well defined helicity, preserved in forward scattering, $\sigma =1$, impinging on a example sphere with $g = - 0.4$. The scattered light is shown via the conical trajectories of the Poynting vector. At $\theta = \pi/2$, the single measurement of the DoCP gives the value of the scatterer's $g$-parameter. Red and blue lines illustrate both the counterclockwise and clockwise polarizations, while the linear polarization (LP) is illustrated in green. }\label{1} 
\end{figure}
The electric field scattered by the nanosphere can be expanded in the helicity basis, allowing us to separate it into two components with well-defined helicity, ${\bf{E}}_\sigma^{\rm{scat}} = {\bf{E}}_{\sigma +} + {\bf{E}}_{\sigma -}$. 

\begin{figure}
\centering
\fbox{
\includegraphics[width=0.9\columnwidth]{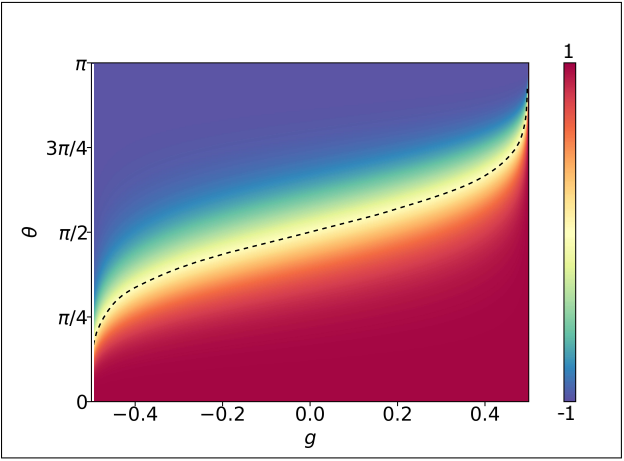}
}
\caption{Color map of the DoCP vs the  scattering angle $\theta$ and the g-parameter. The white vertical line  indicates that this set of DoCP values are forbidden due to causality, i.e., $g > -1/2$. The first Kerker condition, satisfied for $g = 1/2$, gives raise to the conservation of the DoCP, independently of the scattering angle (intense red-color). The dashed line illustrates the curve where the scattered light is linearly  polarized, $\Lambda = 0$.  }
\label{2} 
\end{figure}

For subwavelength spheres, characterized by their first $a_1$ and $b_1$ Mie coefficients \cite{bohren2008absorption} or by their
electric and magnetic polarizabilities, $\alpha_{\rm{E}} = i a_1 k^3 / 6 \pi$ and $\alpha_{\rm{M}} = i b_1 k^3 / 6 \pi$,  the (FF) scattered fields are given by  \cite{olmos2018enhanced}
\begin{equation}\label{SO:2}
{\bf{E}}_{\sigma \sigma'} \sim \frac{1}{\sqrt{2}} E_{\sigma\sigma'} e^{ i \sigma \varphi} ( {\bf{\hat{e}}}_\theta + i \sigma' {\bf{\hat{e}}}_\varphi) , 
\end{equation}
\begin{equation} \label{SO:3}
\frac{E_{\sigma \sigma'}}{E_0} = \frac{e^{ikr}}{4\pi kr}  k^3 \left( \frac{\sigma\alpha_{\rm{E}} + \sigma' \alpha_{\rm{M}}}{2} \right) (\sigma\cos\theta + \sigma' ) ,
\end{equation}
where $E_0$ is the amplitude of the incident wave. The (FF) radiation pattern, given by the differential scattering cross section, is given by \begin{equation}
\frac{d \sigma_{\text{scatt}}}{d \Omega} =\frac{k^4 \left(|\alpha_{\rm{E}}|^2 +  |\alpha_{\rm{M}}|^2\right) }{32 \pi^2}   \left( {1 + \cos^2 \theta}  +4 g  \cos \theta \right) \label{FFI}
\end{equation}
where 
 \begin{equation}\label{SO:5}
g = \langle cos \theta \rangle  = \frac{\text{Re}\left\{  \alpha_{\rm{E}}  \alpha^*_{\rm{M}}\right\}}{|\alpha_{\rm{E}}|^2 +  |\alpha_{\rm{M}}|^2} 
\end{equation}
is the asymmetry parameter in the dipolar approximation \cite{gomez2012negative}.  Notice that $-1/2 < g \leq 1/2$  where the limits correspond to the so-called first ($g=1/2$) and second ($g=-1/2$ ) Kerker conditions \cite{kerker1983electromagnetic}, being $g=-1/2$ an unreachable value in the absence of gain \cite{alu2010does,gomez2011electric}.

\begin{figure*}[t ]
\centering
\fbox{
\includegraphics[width=2\columnwidth]{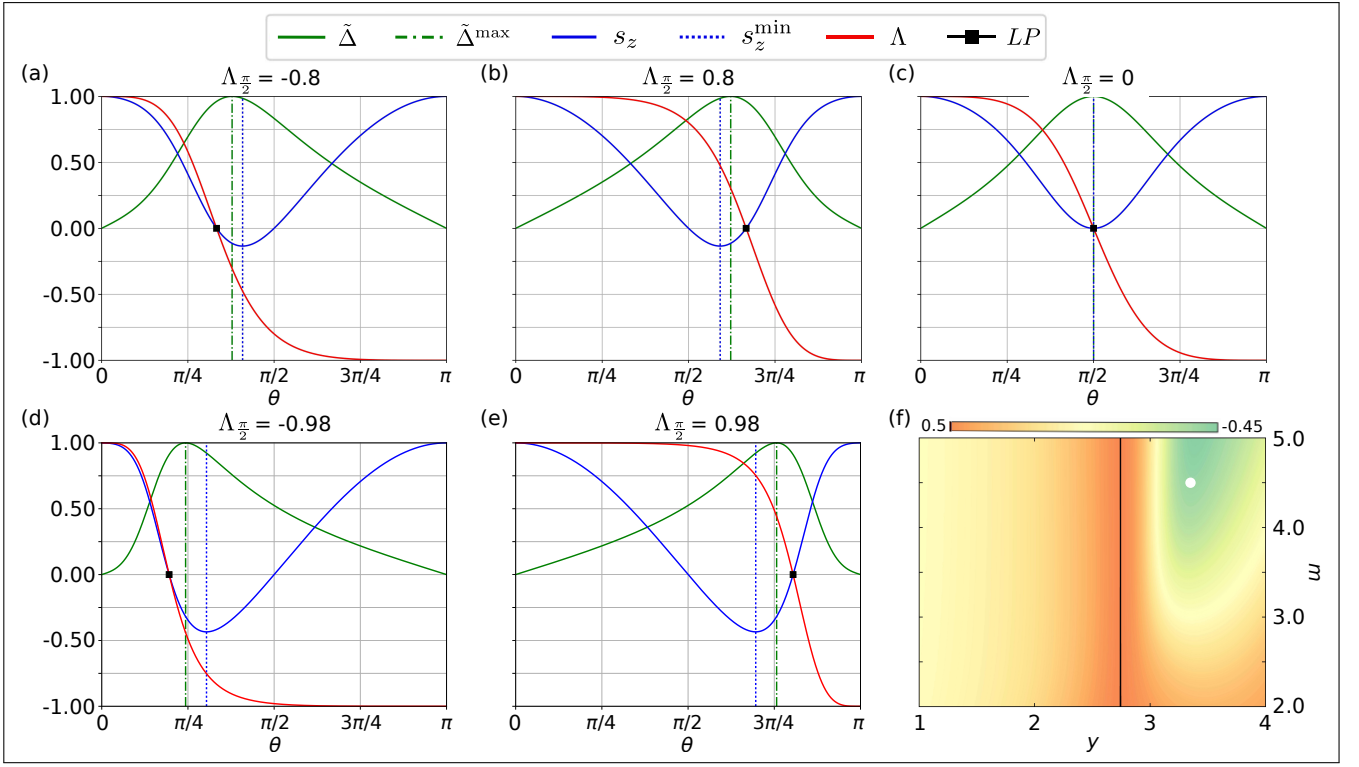}
}
\caption{Normalized  optical mirage $\tilde{\Delta} = \Delta  / \tilde{\Delta}^{\rm{max}}$, spin density ($s_z$) and DoCP ($\Lambda$) vs the scattering angle $\theta$ for an incoming circularly polarized plane wave of helicity $\sigma=+1$. The green  vertical dashed-dotted lines represent the angles corresponding  to the maximum of optical mirage, $\tilde{\Delta}^{\rm{max}}$. Blue  dotted lines correspond to the angles at which  spin-to-orbit angular momentum transfer is maximum (or minimum value of the $z$-component of SAM per scattered photon, $s^{\rm{min}}_z$). The black squares indicate $\Lambda_\theta = 0$, namely, the (FF) observation angles at  which light is linearly polarized (LP). (a) and (b) correspond to $\Lambda_{\pi/2} = -0.8$ and  $\Lambda_{\pi/2} = +0.8$ (i.e. $g =-0.4$ and $g =+0.4$), respectively.   As it can be seen, for $g \neq 0$, $\tilde{\Delta}^{\rm{max}}$, $s^{\rm{min}}_z$ and LP are localized in three different scattering angles, In contrast with the  $g=0$ case (c)  where they all  collapse at right scattering angles $\theta = \pi /2$. 
(d) and (e) show the different angular dependences as the asymmetry parameter approaches the second ($\Lambda_{\pi/2} = -0.98 \gtrsim -1$) and first  ($\Lambda_{\pi/2} = 0.98 \lesssim 1$) Kerker conditions, respectivelly.
(f) reproduces the asymmetry parameter for isotropic spheres as a function of their refractive index $m$ and size parameter $y= mka$ in the dipolar regime (after Ref. \cite{gomez2012negative}). Black vertical line indicates the first Kerker condition, where $\alpha_{\rm{E}} = \alpha_{\rm{M}}$. The solid white point highlights $g =-0.4$ which corresponds to both a high refractive index (HRI) dielectric sphere or to a small perfectly conducting sphere \citep{jackson1999electrodynamics}. Both subwavelength particles will give rise to exactly the same spin-orbit coupling effects.    }\label{3} 
\end{figure*}

Appliying the definition of the helicity operator \cite{fernandez2012helicity}, $\boldsymbol{\Lambda} \equiv (1/k)  \boldsymbol{\nabla} \times$,  the helicity density or DoCP can be expressed in terms of the $V$ and $I$  Stokes parameters
\begin{align}\label{SO:1}
\text{DoCP} =\Lambda_\theta &= \frac{{{\bf{E}}_\sigma^{\rm{scat*}}} \cdot \left(  \boldsymbol{\Lambda} {\bf{E}}^{\rm{scat}}_\sigma \right)}{{{\bf{E}}_\sigma^{\rm{scat}}}^* \cdot {\bf{E}}^{\rm{scat}}_\sigma} = \frac{|{\bf{E}}_{\sigma +}|^2 - |{\bf{E}}_{\sigma -}|^2}{|{\bf{E}}_{\sigma +}|^2 + |{\bf{E}}_{\sigma -}|^2} = \frac{V}{I} \\
&=  \frac{ 2  \sigma \left(\left(1 + \cos^2 \theta  \right)g  +  \cos \theta \right)}{{1 + \cos^2 \theta}  +4 g  \cos \theta }, \label{SO:1b}
\end{align}
while  the DoCP mean value $\langle \Lambda \rangle$ \citep{schwartz2006conservation}
\begin{equation}
\langle \Lambda \rangle \equiv \frac{\int \left\{ |{\bf{E}}_{\sigma +}|^2 - |{\bf{E}}_{\sigma -}|^2\right\} d\Omega }
{\int \left\{ |{\bf{E}}_{\sigma +}|^2 + |{\bf{E}}_{\sigma -}|^2\right\} d\Omega} = 2\sigma g .
\end{equation}
The angular dependence of the DoCP  just depends on the $g$-parameter. In Fig. \ref{2} we show the DoCP pattern vs. both scattering angle $\theta$ and $g$-parameter for an incoming light with helicity $\sigma=+1$. As it can be inferred, the DoCP values are restricted to $-1 < \Lambda \leq 1$, being maximized when the system is dual, i.e., at the first Kerker condition when helicity is preserved. In addition, we find that the polarization of the scattered light is linear ($\Lambda_{\theta_0} = 0$) when the condition $g = - \cos \theta_0 / (1+ \cos^2 \theta_0)$ is fulfilled, corresponding to the dashed line in Fig. \ref{2}. As it can be seen, it matches with $\theta_0 = \pi/2$ only for $g=0$, which corresponds with the pure electric (or magnetic) dipolar case. 
The relatively simple  measurement of the polarization degree at a right-angle scattering configuration provides a useful insight on the scattering properties of small particles. In particular, the spectral evolution of the degree of linear polarization  was shown to be a simple and accurate way to identify electric and magnetic behaviours of the scattered fields \cite{geffrin2012magnetic,setien2010spectral,garcia2010linear}.
Interestingly,  we find that the DoCP  measured  at  a right scattering angle, $\theta = \pi/2$,  is directly  related to the parameter $g$:
\begin{equation}\label{SO:6}
\Lambda_{\frac{\pi}{2}} = \langle \Lambda \rangle = 2\sigma g \; .
\end{equation}
This is one of the  important results of the present work, which states  
that by measuring the degree of circular polarization at $90^o$ degrees, we can directly extract the  $g$-parameter. 


Once we have a complete description of the angular dependence of the helicity density in the dipolar regime, it is interesting to  analyze its relation with the angular momentum exchanges and the spin-orbit optical mirage. Following Crichton and Marston \citep{crichton2000measurable}, we notice that the $z$-component of the SAM  per scattered photon, $s_z(\theta)$, is a measurable quantity simply related to the  DoCP of the scattered light  \begin{equation} s_{z} = \Lambda_\theta \cos \theta, \label{sz} \end{equation} 
where $ \Lambda_\theta$ is given by  \eqref{SO:1b}.
Additionally, due to the axial symmetry of the scatterer, the $z$-component of the {\em total} angular momentum of the incoming photons $j_z = \sigma$ is preserved after scattering. Then the $z$-component of the OAM  per scatterd photon, $\ell_z(\theta)$,  can also be related to the DoCP  \begin{equation} \ell_{z}(\theta) \equiv j_z -s_z(\theta) = \sigma - \Lambda_\theta \cos \theta, \label{lz} \end{equation}  which allow us to link the optical mirage's apparent shift \cite{olmos2018enhanced}, $\Delta$, with  $ \Lambda_\theta$ as 
\begin{align}\label{SO:7}
\frac{\Delta \pi}{\lambda} =   \frac{\ell(\theta)}{\sin\theta} = \frac{\sigma - \Lambda_\theta \cos\theta}{\sin\theta} .
\end{align}

Equations \ref{FFI}, \ref{SO:1b}, \ref{sz}, \ref{lz} and  \ref{SO:7} demonstrate that observables as the intensity, degree of circular polarization (DoCP), spin and orbital angular momentum of scattered photons and the optical mirage   ($d\sigma_\text{scatt}/d \Omega$, $\Lambda_\theta$, $s_z$, $\ell_z$ and $\Delta$) 
are all linked, and fully determined, by  the dimensionless  ``asymmetry parameter'' $g$, being independent of  the specific optical properties of the scatterer.  In other words,  within a single measurement of the DoCP at $90^o$ via polarization filters in the FF, we can extract  the $g$-parameter via Eq. \ref{SO:6}, and infer the angular dependence of all the relevant scattering quantities. 

Figure 3 illustrates the angular momentum exchange and the optical mirage dependence with the (FF) observation angle $\theta$ for an incoming plane wave with helicity $\sigma=+1$ and total $z$-component of the total angular momentum per photon $j_z=\sigma$. Figures \ref{3}(a) and  \ref{3}(b) summarize the results  for  $g = -0.4$ and $g = 0.4$, respectively. In contrast with pure electric (or pure magnetic) dipolar particles with symmetric scattering ($g=0$, Fig. \ref{3}(c) ),  the maximum angular momentum exchange (corresponding to the minimum of $s_z$) and the maximum apparent shift of the optical mirage, $\tilde{\Delta} = \Delta / \Delta_{\rm{max}} $,  take place at different scattering angles but in an angular region in which the $z$-component of the SAM is negative (i.e. where the photons are not linearly polarized) while  the $z$-component of the OAM is larger than that of the total AM ($\ell_z = j_z- s_z > j_z = 1$). The equivalent effect occurs for $\sigma=-1$.

The angular gaps between the minimum of the $z$-component of the SAM (maximum AM exchange), the maximum of the optical mirage effect and the angle at which  light is linearly polarized first increase when the asymmetry parameter tends to the second or first Kerker conditions, $|g| \approx 0.5$, as it can be seen in FIGs. \ref{3}(d) and  \ref{3}(e).  However, in the limit of  dual scatterers ($g=+0.5$), we find that  $s_z \rightarrow \cos\theta$ and $\Lambda_\theta \rightarrow +1$, and  
the extrema collapse again at the singular backscattering angle $\theta = \pi$.   At this angle the optical mirage diverges and an optical vortex appears  with $s_z=-1$ and $\ell_z =2$ \citep{olmos2018enhanced}. In contrast, in absence of gain, the Optical Theorem imposes  that the limit of $g = -0.5$ is unreachable \cite{alu2010does,gomez2011electric} which inhibits the  complete (flipping) transformation from $s_z = \sigma$ to $s_z = -\sigma$, although a huge enhancement of the optical mirage is predictable  getting close to this condition.  
As a consequence,  in analogy with  dual spheres, we  can predict that  an anti-dual sphere   -that could be made with a material with gain- \cite{zambrana2013duality}, $g=-0.5$, will generate a perfect optical vortex in the forward direction with a divergent apparent displacement.

In conclusion, we have shown that the asymmetry and spin-orbit coupling effects  of light scattered from subwavelength  spheres with electric and magnetic dipolar responses  are   fully determined by  the dimensionless  ``asymmetry parameter'' $g$. As a consequence, 
particles with different optical properties, but sharing an identical $g$ parameter value (see Fig. \ref{3}(f) ) will lead to the same angular dependences of the intensity, DoCP, SAM to OAM exchanges and optical mirage apparent shifts. 
Since the parameter $g$ can be obtained from a far-field measurement of the DoCP our results  predicts the possibility  of determining all angular dependences of the scattering coefficients  from a single polarization measurement, and therefore
they  open new perspectives in different areas of Optics and Photonics including 
 antennas engineering, metamaterials, nanophotonics and optical imaging.   

This research was supported by the Spanish Ministerio
de Econom\'ia y Competitividad (MINECO, MICINN) and European
Regional Development Fund (ERDF) Projects FIS2014-55987-P, 
FIS2015-69295-C3-3-P and FIS2017-82804-P and by the Basque Dep. de Educación Project PI-2016-1-0041.

\bibliography{SOCLight}


\end{document}